\documentclass[a4paper,fleqn,usenatbib]{mnras}

\usepackage{newtxtext,newtxmath}

\usepackage[T1]{fontenc}
\usepackage{ae,aecompl}

\usepackage{graphicx}	
\usepackage{amsmath}	
\usepackage{amssymb}	

\usepackage{hyperref}
\hypersetup{
  citecolor=cyan,      
}

\usepackage{graphicx}
\usepackage{color}
\definecolor{orange}{rgb}{1,0.5,0}

\newcommand{\Msun}{M_{\odot}}

\newcommand{\MTOV}{M_{_{\mathrm{TOV}}}}

\newcommand{\ie}{i.e.,~}
\newcommand{\eg}{e.g.,~}

\usepackage{amsmath}
\usepackage{amssymb}
\usepackage{mathtools}

\title[A lower bound on the maximum mass from GW190814]{A lower bound on
  the maximum mass if the secondary in GW190814 was once a rapidly
  spinning neutron star}

\author[E.R. Most et al.]{ Elias R. Most$^{1}$, L. Jens Papenfort$^{1}$,
Lukas R. Weih$^{1}$, Luciano Rezzolla $^{1,2,3}$\\ 
$^{1}$ Institut f\"ur Theoretische Physik, Goethe Universit\"at,
Max-von-Laue-Str. 1, 60438 Frankfurt am Main, Germany\\
$^{2}$School of Mathematics, Trinity College, Dublin 2, Ireland\\
$^{3}$Helmholtz Research Academy Hesse for FAIR,
Max-von-Laue-Str. 12, 60438 Frankfurt am Main, Germany}

\date{Accepted XXX. Received YYY; in original form ZZZ}

\pubyear{2020}

\begin{document}
\label{firstpage}
\maketitle

\begin{abstract}
 The recent detection of GW190814 featured the merger of a binary with a
 primary having a mass of $\sim 23\,M_{\odot}$ and a secondary with a
 mass of $\sim 2.6\,M_{\odot}$. While the primary was most likely a black
 hole, the secondary could be interpreted as either the lightest black
 hole or the most massive neutron star ever observed, but also as the
 indication of a novel class of exotic compact objects. We here argue
 that although the secondary in GW190814 is most likely a black hole at
 merger, it needs not be an ab-initio black hole nor an exotic
 object. Rather, based on our current understanding of the nuclear-matter
 equation of state, it can be a rapidly rotating neutron star that
 collapsed to a rotating black hole at some point before merger. Using
 universal relations connecting the masses and spins of uniformly
 rotating neutron stars, we estimate the spin, $0.49_{-0.05}^{+0.08}
 \lesssim \chi \lesssim 0.68_{-0.05}^{+0.11}$, of the secondary -- a
 quantity not constrained so far by the detection -- and a novel strict
 lower bound on the maximum mass, $\MTOV > 2.08^{+0.04}_{-0.04}\,
 \,M_{\odot}$ and an optimal bound of $ \MTOV > 2.15^{+0.04}_{-0.04}\,
 \,M_{\odot}$, of nonrotating neutron stars, consistent with recent
 observations of a very massive pulsar. The new lower bound also remains
 valid even in the less likely scenario in which the secondary neutron
 star never collapsed to a black hole.
\end{abstract}

\begin{keywords}
transients: black hole - neutron star mergers --- gravitational waves ---stars: neutron
\end{keywords}

\section{Introduction}
\label{sec:intro}

With the detection of gravitational waves (GW) from a binary black hole
(BBH) merger GW190412 \citep{Abbott2020a} with mass ratio of $q =
0.28_{-0.06}^{+0.13}$ by the LIGO/Virgo collaboration, it has been shown
that even for more massive BBH a significant mass ratio can be
acquired. While the lower mass companion with a mass of $m_2 =
8.4_{-1.0}^{+1.8}\,M_{\odot}$ is definitely a black hole (BH) in this
case, it gave an interesting prospect for the possibility of even more
asymmetric binary mergers in the future.

The most recent detection GW190814 \citep{Abbott2020b} belongs to a
binary merger featuring a much lower mass ratio of $q =
0.112_{-0.009}^{+0.008}$ with a massive primary companion of $m_1 =
23.2_{-1.0}^{+1.1}\,M_{\odot}$ and a secondary of $m_2 =
2.59_{-0.09}^{+0.08}\,M_{\odot}$, which falls in the possible mass gap
between neutron stars (NSs) and stellar BHs.
Being the source of the most asymmetric binary compact object merger to
date GW190814 seems to challenge the established binary formation
channels. It is argued in \citet{Abbott2020b} that the formation of such
a high mass ratio system as an isolated binary is strongly suppressed in
population synthesis simulations. This is certainly true at Milky-Way
metallicity where stellar winds are much stronger and progenitors lose
more mass throughout the binary evolution. This leads to lower maximum BH
masses and subsequently limits the maximum-mass ratio for BBH as well as
BH-NS systems \citep{Kruckow2018}. However, this already posed a challenge
for previous detection of BBH with large total mass. It is assumed that
these systems have to originate from a lower metallicity environment
\citep[see, \eg][]{Stevenson2017a}. Especially \citet{Kruckow2018} have
shown that there is a significant population of BBH as well as BH-NS
systems with low mass ratios at low metallicity which match the source of
GW190814. Together with the broad range of resulting merger times the
birth as an isolated binary constitutes still a viable formation channel,
although it challenges current supernova explosion mechanisms
\citep{Zevin2020}. Alternative formation scenarios might include the merger
of quadruple systems \citep{Hamers2020} or circumbinary accretion in a
highly asymmetric progenitor system \citep{Safarzadeh2020b}. 
Both, BBH and BH-NS systems, with strong asymmetry are nonetheless
suppressed compared to equal mass BBH binaries of the same total mass,
due to mass transfer during the binary evolution equalising the companion
masses to less asymmetric configurations. Additional alternative channels
through dynamical captures or hierarchical binary systems are presented
in \citet{Abbott2020b}. While the formation rates through these processes
are not well known, they all need dense stellar environments to work and
the probability to form such a system is independent of the low mass
companion being a BH or a NS in these cases.
Assuming a mass gap between NSs and BHs at approximately $~5\Msun$, the
progenitor of the secondary companion has to be a NS. For both
possibilities of forming the system in isolation or through dynamical
processes, accretion of significant amounts of matter onto the NS or
gaining angular momentum afterwards is very unlikely. Either due to the
fact that the primary companion is evolving faster and thus collapses to
a BH before the NS is formed, or because formation in dense stellar
systems is only significant in mass segregated regions for which the
timescale is much larger than the stellar evolution up to formation of
the compact object.

Combining these implications, we conclude that the NS companion has to be
born already with an approximate mass of $2.59\,M_{\odot}$ which is above
the upper limit for the maximum mass for a nonrotating NS, $2.33\,
M_{\odot}$ \citep{Rezzolla2017,Shibata2019}. Consequently, such a NS has
to be supported by rotation against gravitational collapse for a
significant time after its birth. Long-term electromagnetic spin-down
might lead to its collapse to a BH if a significant fraction of the spin
has been removed.  Such a BH would then inherit its mass and spin from
the NS.

For the rest of the paper, we will assume that the secondary in GW190814,
hence, was either a rapidly rotating NS or a BH formed by the
gravitational collapse of such with the same properties.  Given the well
constrained mass of the low mass companion, we are able to use the
universal relations between maximum supportable mass $M_{\rm crit}$ and
the ratio of the stars dimensionless spin to its maximum dimensionless
spin at the mass shedding limit found by \cite{Breu2016} to find a new
lower bound on the maximum mass $\MTOV$ a nonrotating NS must be able to
support.
Such a constraint can help to constrain the equation of state (EOS) of
nuclear matter at densities beyond the reach of earth-based experiments.
Indeed, in the recent past a number of studies have used the
multimessenger signal of the event GW170817 \citep{Abbott2017} in order
to derive astrophysical constraints and/or translate them into
constraints on the EOS \citep{Margalit2017,
  Bauswein2017b,Rezzolla2017,Ruiz2017,Annala2017,Radice2017b,Most2018,De2018,
  Abbott2018b,Montana2018,Raithel2018,Tews2018,Malik2018,Koeppel2019,Shibata2019}.
This work falls in line with these studies by deriving a new lower limit
on $\MTOV$.
Additionally, we give a lower bound on the dimensionless spin for the
secondary companion, using the upper maximum-mass constraints from the
GW170817 event \citep{Rezzolla2017, Shibata2019}.

\section{Rapidly spinning neutron stars: The basic picture}
\label{sec:method}

A binary system of two compact objects can be described in terms of their
masses $m_1$ and $m_2\, \, \left( < m_1 \right)$ and corresponding
dimensionless spins ${\chi}_1 = S_1/ m_1^2$ and ${\chi}_2 = S_2/m_2^2$,
where $S_1$ and $S_2$ are the spin angular momentum of the two compact
objects. For simplicity, we will consider only the component of the spins
aligned with the orbital angular momentum, which is usually extracted
from gravitational wave observations \citep{Abbott2020b}. The effective
spin is defined as
\begin{equation}
  \label{eqn:chitilde}
  \tilde{\chi}
  :=\frac{m_1 \chi_1 + m_2 \chi_2 }{m_2 + m_1} = \frac{\chi_1}{1+q}\left( 1
  + q\frac{\chi_2}{\chi_1} \right),
\end{equation}
where $q := m_2/m_1 \leq 1$ is the mass ratio of the binary system.

If one of the objects in the system is a NS, its maximally allowed mass
$M_{\rm crit}$ will depend on its spin $\chi_1$. In particular, for a
nonrotating NS $m_1 < \MTOV$, where $\MTOV$ is the maximum mass of a
nonrotating NS. Observations of PSR J0348+0432 indicate that $\MTOV >
2.01^{+0.04}_{-0.04}\,M_{\odot}$ \citep{Antoniadis2013}, with a recent
observation of PSR J0740+6620 indicating that $\MTOV >
2.14^{+0.10}_{-0.09}\,M_{\odot}$, although with this large uncertainty
already at the $1\rm{-}\sigma$ level \citep{Cromartie2019}. While the GW
signal of GW170817 did not rule out EOSs consistent with nuclear
constraints and small maximum masses, $\MTOV \simeq 2.0 M_\odot$
\citep{Malik2019}, the constraints imposed by the electromagnetic
counterparts of GW170817 \citep{Abbott2017_etal,Abbott2017b} have led to
upper bounds on the maximum mass of $\MTOV \lesssim 2.3\,\Msun$
\citep[see, \eg][]{Margalit2017,Rezzolla2017,Ruiz2017,Shibata2019},
although this value could be higher in the less likely case that GW170817
did not result in a BH \citep{Ai2019}. We discuss the impact of the
assumption of an upper maximum mass at the end of the paper.

\noindent If the NS is spinning, its maximum mass can be higher than
$\MTOV$ due to the additional rotational support against gravitational
collapse to a BH. In particular, \citet{Breu2016} have found that the
critical mass $M_{\rm crit}$, that is, the mass of uniformly rotating NSs
on the mass stability line, can be expressed in a (quasi-) universal
relation through the dimensionless spin on the stability line $\chi_{\rm
  crit}$ \citep{Friedman88,Takami:2011}, and the maximum dimensionless
spin at the mass shedding limit $\chi_{\rm kep}$\footnote{Note that
  maximum mass supported through uniform rotation is not the end-point of
  neither the turning-point line \citep{Friedman88}, nor of the
  neutral-stability line \citep{Takami:2011}, and it is always larger
  than both of them \cite[see, \eg][for details]{Weih2017,Bozzola2019}.}
\begin{align}
  M_{\rm crit}(&\chi_{\rm
  crit}, \chi_\mathrm{Kep}, M_{_{\rm TOV}}) := \nonumber\\ &M_{_{\rm TOV}} \left( 1 +
  a_1 \left( \frac{\chi_{\rm{crit}}}{\chi_{\rm Kep}} \right)^2 +a_2 \left(
\frac{\chi_{\rm{crit}}}{\chi_{\rm Kep}} \right)^4 \right)\,,
\label{eqn:M_crit}
\end{align}
where $a_1 = 0.132$, $a_2 = 0.071$. This directly implies that the
maximum mass for any uniformly rotating NS is limited by the spin at
the mass-shedding limit, $\chi_{\rm crit} = \chi_{\rm Kep}$, where
\begin{align}
M_{\rm max}:=M_{\rm crit}(\chi_\mathrm{crit} = \chi_{\rm{Kep}})=\xi_{\rm
max}\,M_{_{\rm TOV}} ~, ~ \xi_{\rm max} = 1.203\pm0.022\,,
\label{eqn:MmaxXimax}
\end{align}
as was shown in \citep{Breu2016}. Note that this result, combined with the present
estimates on $\MTOV$ implies a \textit{strict} upper limit on the mass of a NS
supported via uniform rotation, \ie $M_{\rm max}\leq 2.85\,M_{\odot}$,
which is clearly larger than the mass inferred for the secondary in
GW190814.

\begin{figure}
  \centering
  \includegraphics[width=0.40\textwidth]{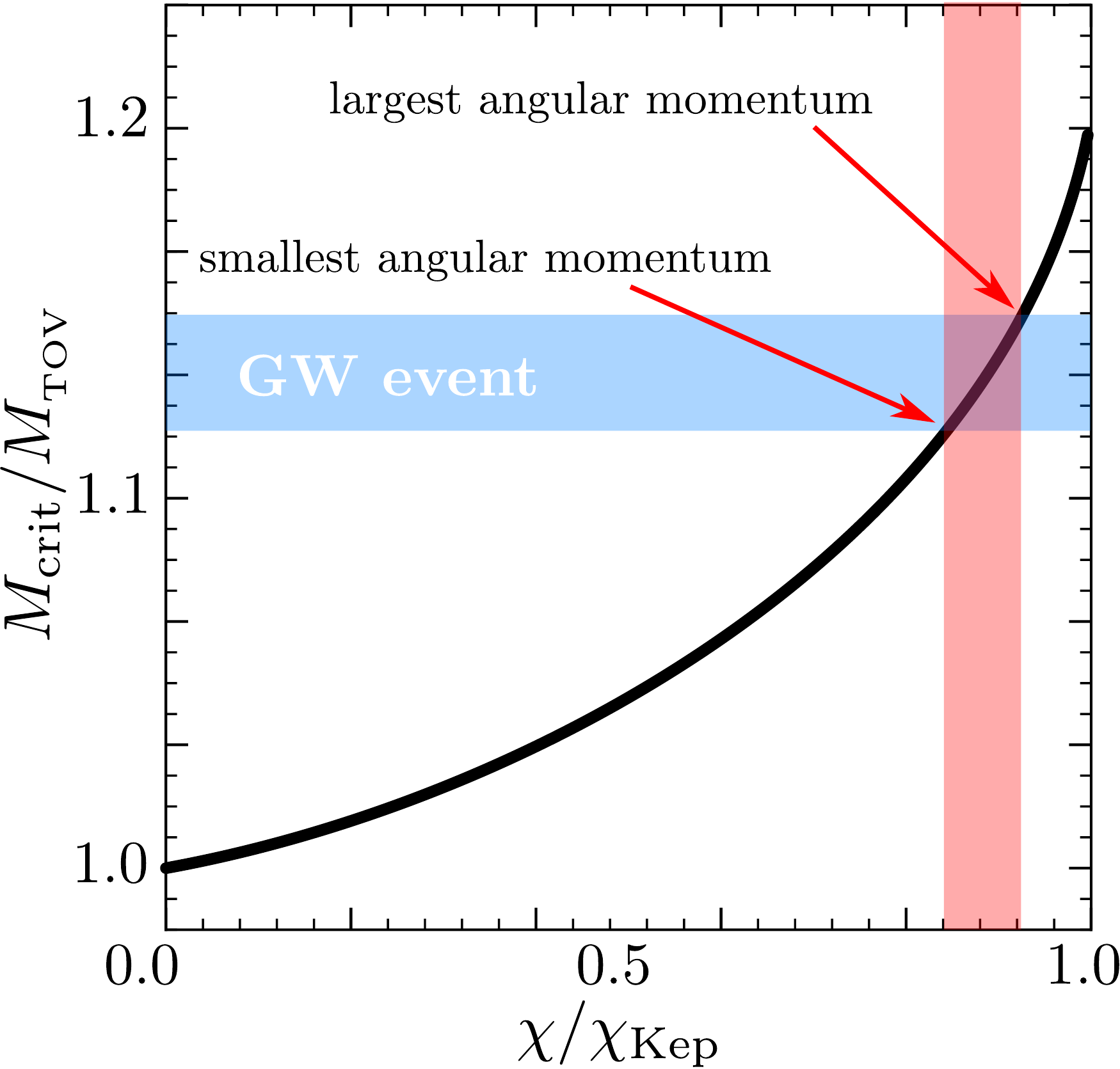}
  \caption{Schematic representation of the universal relation for
    uniformly rotating NSs along the stability line,
    \citep{Breu2016}. The solid black line reports the (quasi-)universal
    relation between $M_{\mathrm{crit}}/\MTOV$ and the dimensionless
    angular momentum of the star when normalised to its maximum value
    $\chi/\chi_{\mathrm{Kep}}$, while the blue-shaded area are the
    constraints on the mass from a given GW event. For any chosen
    $\MTOV$, the intersections of the blue-shaded area with the solid
    line will select the rotating stars having the smallest and the
    largest angular momentum. }
  \label{fig:sketch}
\end{figure}

\noindent Given a mass $M_{\rm crit}$ of a rapidly rotating NS it is possible to
set bounds on its spin $\chi$ and the maximum mass $\MTOV$ of a
nonrotating NS. This is shown schematically in Fig. \ref{fig:sketch},
which reports the universal relation for the masses of NSs along the
stability line of uniformly rotating models $M_{\mathrm{crit}}$
\citep{Breu2016}. More specifically, the solid black line reports the
(quasi-)universal relation between $M_{\mathrm{crit}}/\MTOV$ and the
dimensionless angular momentum of the star when normalised to its maximum
value $\chi/\chi_{\mathrm{Kep}}$, \ie Eq. (\ref{eqn:M_crit}). One can
see, for instance, that the largest possible mass for a rotating star
$M_{\mathrm{max}}$ is $\sim 1.2$ times that of the corresponding
nonrotating model for any EOS. Shown instead with a blue-shaded area are
the constraints on the mass from a given GW event. This area will depend
not only on the GW measurement, but also on $\MTOV$. The intersections of
the blue-shaded area with the solid line will then select the rotating
star on the stability limit having the smallest and the largest angular
momentum that is still in agreement with the observation.  The red-shaded
area will therefore represent the allowed range in spin for a massive NS
that has collapsed to a BH.
Since $\chi \leq \chi_{\mathrm{Kep}}$, it is also possible to determine a
lower bound on $\MTOV$ when $M_{\mathrm{crit}}=M_{\mathrm{max}}$, that
is, when the lower limit of the observed mass range (lower edge of
blue-shaded area) is reached by a star with $\chi=\chi_{\mathrm{Kep}}$.

While the maximum mass $M_{\rm max}$ supported by a rapidly rotating
NS does not depend on the numerical value of $\chi_{\rm Kep}$, it will turn
out to be useful to give a more precise value that will be necessary to
constrain the spin of the secondary in GW190814.
Based on the results of \citet{Koliogiannis2020}, it is possible to
express $\chi_{\rm Kep}$ in terms of the compactness $\mathcal{C}_{\rm
  TOV} = \MTOV/R_{_{\mathrm{TOV}}}$, \ie \citep{Breu2016,Shao2020}
\begin{align} \chi_{\rm
    Kep} \simeq \frac{\alpha_1}{\sqrt{\mathcal{C}_{_{\rm TOV}}}} +
  \alpha_2 \sqrt{\mathcal{C}_{_{\rm TOV}}}\, . \label{eqn:chik}
\end{align}
where $\alpha_1= 0.045 \pm 0.021$ and $\alpha_2= 1.112 \pm 0.072$
\citep{Most2020c}. Using a large set of EOSs compatible with current
bounds on the tidal deformability and constraints on the maximum mass of
nonrotating NSs from GW170817 \citep{Most2018,Weih2019}, it was possible
to show that a lower/upper bound in compactness can be given in terms of
$M_{_{\rm TOV}}$ \citep{Most2020c}, \ie
\begin{eqnarray}
  \label{eqn:C_TOV_max}
  & \sqrt{\mathcal{C}_{_{\rm TOV}}^{\rm min}} =
  c_1 M_{_{\rm TOV}} +
  c_2 M^2_{_{\rm TOV}} +
  c_3 M^3_{_{\rm TOV}}\,,\\
  & \sqrt{\mathcal{C}_{_{\rm TOV}}^{\rm max}} =
  d_1 M_{_{\rm TOV}} +
  d_2 M^2_{_{\rm TOV}} +
  d_3 M^3_{_{\rm TOV}}\,,
  \label{eqn:C_TOV_min}
\end{eqnarray}
where $c_1=0.482$, $c_2=-0.174$, $c_3 = 0.027$, $d_1=0.6522$,
$d_2=-0.256$, $d_3 = 0.034$.
\section{Application to GW190814}
Recently the LIGO/Virgo collaboration has reported the detection of the
merger of a $\sim 23\,\Msun$ BH with another $\sim 2.6\,M_{\odot}$
compact object \citep{Abbott2020b}. Following the scenarios outlined in
Sec. \ref{sec:intro} we assume that we have the merger of a very
massive BH with a light BH that was produced by the collapse of a rapidly
spinning NS prior to merger. We can then use the universal relations summarised in
Sec. \ref{sec:method} to extract bounds on the spin $\chi_2$ and $m_2$ of
the secondary companion. Indeed, as we will comment later on, our bounds
apply unchanged even if the rapidly spinning NS never collapsed to a BH.

\begin{figure}
  \centering
  \includegraphics[width=0.37\textwidth]{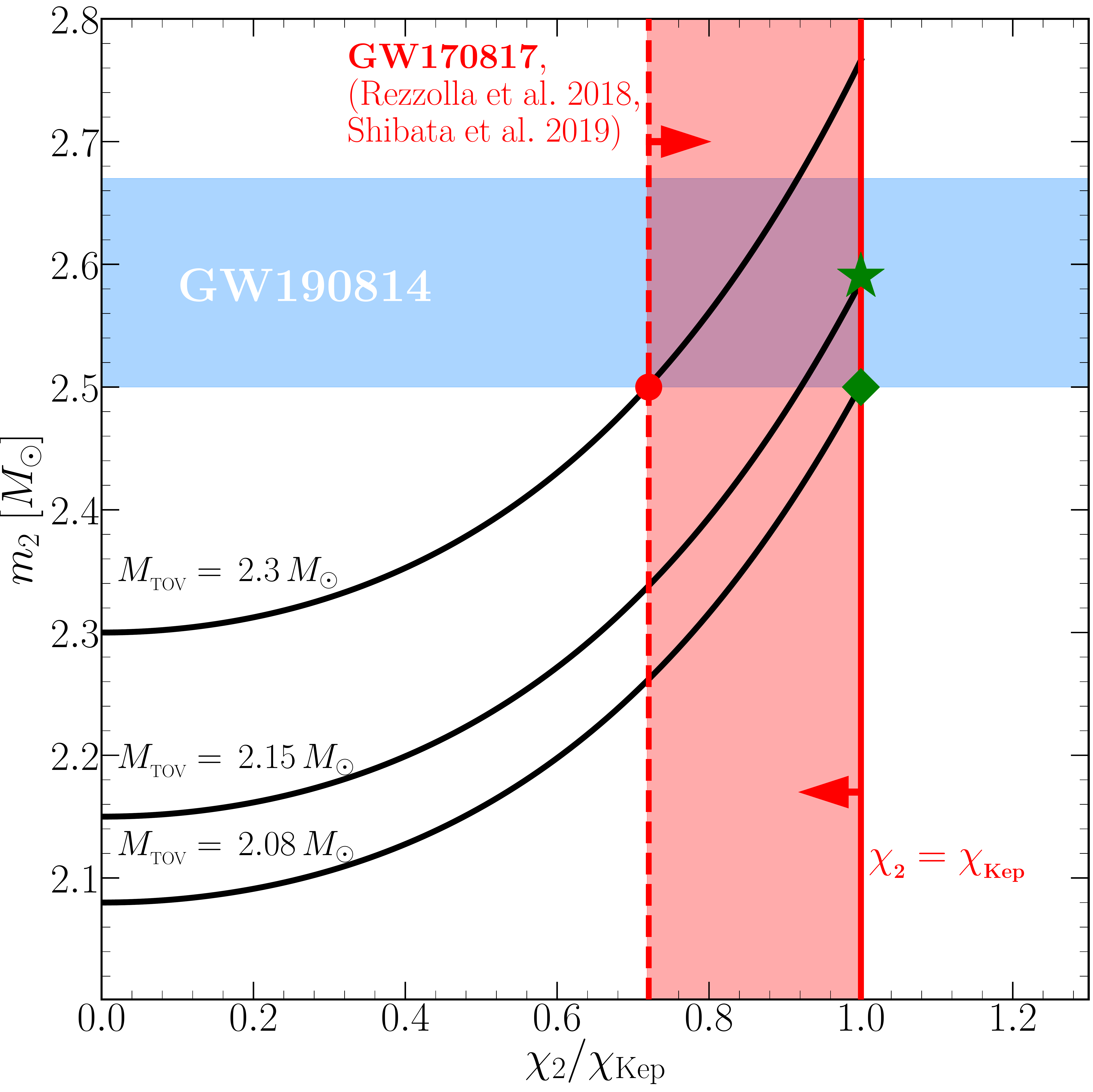}
	\caption{Mass over spin of the secondary in GW190814. The black
	lines show the Eq. (\ref{eqn:M_crit}) assuming that the secondary
	was a critically spinning NS at some point. The red-shaded area
	marks the allowed spin range given the mass measurement (blue
	shaded). The green diamond selects the curve with the minimal
	possible value of $\MTOV$, the green star the one corresponding
      to the most likely inferred mass of GW190814. We note that these
      constraints are independent of the actual value of $\chi_{\rm Kep}$.}
  \label{fig:chi3}
\end{figure}

\noindent In Fig. \ref{fig:chi3} we show the universal relation \eqref{eqn:M_crit}
of the mass $m_2$ of a rapidly spinning NS with its spin $\chi_2$. In
addition, we also shade the allowed region of masses of the secondary from
the GW190814 event in blue \citep{Abbott2020b}, \ie
$m_2=2.59_{-0.08}^{+0.08}$, and mark the maximally allowed spin
$\chi_2=\chi_\mathrm{Kep}$ approximated as a constant with a red vertical
line. Note that for different values for the maximum mass $\MTOV$,
Eq. \eqref{eqn:M_crit}, generates a sequence of rotating stars starting
at $m_2=\MTOV$ and terminating at their maximum value when $\chi_2 =
\chi_{\rm Kep}$. If the secondary binary companion in GW190814 was at
some point a NS (and since then did not change its mass significantly),
it will have to lie on or below one of these sequences. As outlined in
Sec. \ref{sec:method}, the sequence with the lowest $\MTOV$ that still
intersects with the measurement of GW190814 marks (green diamond in
Fig. \ref{fig:chi3}) a lower limit on $\MTOV$, \ie $M_{_{\rm TOV}}^{\rm
  min}$. For the NS to have been stable initially, it could not
have been more massive than the heaviest rotating configuration, \ie
\begin{align}
  m_2 \leq M_{\rm max} \simeq \xi_{\max}\, M_{_{\rm TOV}}^{\rm min}\,,
  \label{eqn:max_m2}
\end{align}
where we have used Eq. \eqref{eqn:MmaxXimax}.
Hence, we find that if the secondary in GW190814 was either a BH formed
by the collapse of a rapidly rotating NS or a stable rapidly rotating NS,
this yields a lower bound on the maximum mass of nonrotating stars, \ie
\begin{align}
	\MTOV \gtrsim M_{_{\rm TOV}}^{\rm min}=m_2^{\rm GW190814}/
	\xi_{\rm max} \approx 2.08 \pm 0.04\,M_{\odot}\,,
  \label{eqn:mtov_lo}
\end{align}
where we have taken the most conservative lower limit on the companion
mass $m_2^{\rm GW190814} = 2.51\,M_{\odot}$ \citep{Abbott2020b}.  For
completeness, we also report the lower bound $\MTOV^{\rm opt} \approx
2.15 \pm 0.04\, \,M_{\odot}$ when using the most likely value of
$m_2^{\rm opt} = 2.59\, M_\odot$.
The corresponding sequence of critical masses of rotating NSs with spin
$\chi_2$ is given by the lower black line in Fig. \ref{fig:chi3}, which
terminates at $\chi_2 = \chi_{\mathrm{Kep}}$.
Two important remarks need to be made at this point as they are sometimes
confused or misunderstood. First, \textit{$M_{\rm crit}$ depends only on
  $M_\mathrm{TOV}$ and not on the actual value of $\chi_\mathrm{Kep}$.}
This is because $M_{\rm crit}$ is deduced from universal relations
expressed in terms of the normalised angular momentum
$\chi/\chi_\mathrm{Kep}$, de-facto removing any information on the
precise (and EOS-dependent) value of $\chi_\mathrm{Kep}$. \textit{Hence,
  the lower maximum-mass bound in Eq. \eqref{eqn:mtov_lo} is universal
  and agnostic of the EOS.}  Second, while the lower bound
\eqref{eqn:mtov_lo} is compatible with the measurement of
\citet{Cromartie2019}, the latter also has a rather large uncertainty,
\ie $2.14^{+0.10}_{-0.09}$ at $1-\sigma$ and $2.14^{+0.20}_{-0.18}$ at
$2-\sigma$. Hence, the bound in Eq. \eqref{eqn:mtov_lo} -- whose
derivation follows a completely different route -- provides
\textit{independent and complementary} strength to the idea that NSs with
masses $\gtrsim 2.1\,M_{\odot}$ should be measured in the near future.

\begin{figure}
  \centering \includegraphics[width=0.37\textwidth]{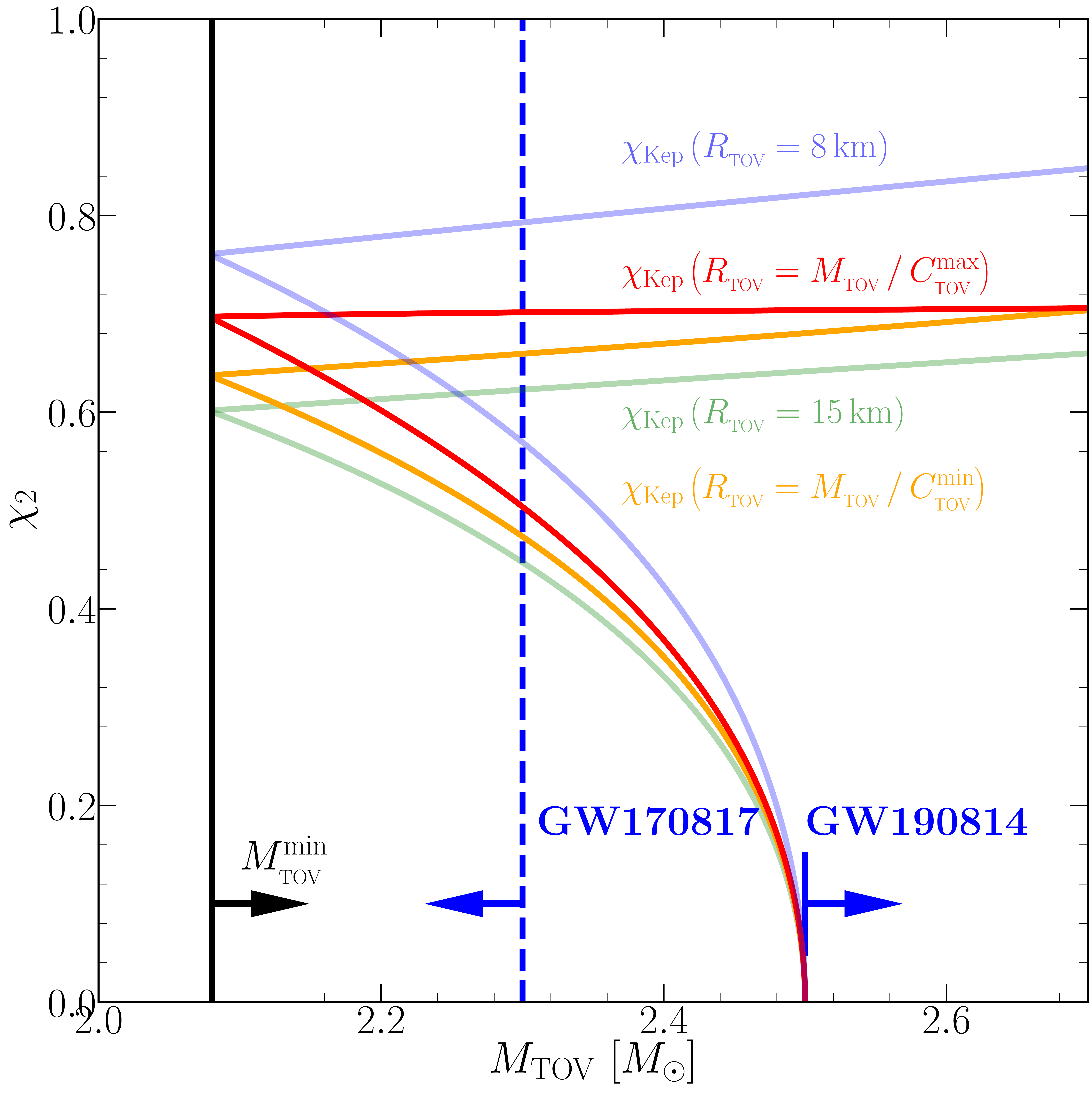}
  \caption{Dimensionless spin bounds for the secondary object. The upper
    (almost constant) lines refer to the Keplerian limit, which
    corresponds to the highest spin a uniformly rotating NS can attain.
    The lower curves correspond to the minimal amount of spin needed to
    support a $2.5\,M_\odot$ star by means of (rapid) rotation. The
    various lines represent different values of $\chi_{\rm Kep}$ as
    described in the figure and the text. The blue lines denote 
    bounds from GW190814 and GW170817, respectively. The limiting spin
    configuration where the two lines intersect, will always happen at
    $M_{_{\rm TOV}}^{\min}$ due to the universal relation between $M_{\rm
      max}$ and $M_{_{\rm TOV}}$.}
  \label{fig:appA}
\end{figure}

\noindent In accordance with the maximum-mass constraints described in
Sec. \ref{sec:method}, we can consider the same logic and draw lines for
any value of $\MTOV \lesssim 2.3\,\Msun$, which is shown with different
black lines in Fig. \ref{fig:chi3}.
From the intersection of the line corresponding to $\MTOV=2.3\,\Msun$
with $m_2^{\rm{GW}190814}=2.51\, \Msun$ we deduce a lower bound of
\begin{align}
   \chi_2  / \chi_{\rm Kep}\, \gtrsim\, 0.72 \,.
  \label{eqn:spin}
\end{align}
In order to translate this constraint into an actual spin constraint, we
need to fix the spin at break-up, $\chi_{\rm Kep}$.
Using Eqs. \eqref{eqn:C_TOV_max} and \eqref{eqn:C_TOV_min} in Eq.
\eqref{eqn:chik} for $\chi_{\rm Kep}$, we find an average
value of $\chi_{\rm Kep}=0.68$ for $2.01 \leq M_{_{\rm TOV}}/M_\odot \leq
2.3$, in accordance with those EOSs. This is consistent with an earlier
constraint of $\chi_{\rm Kep} \simeq 0.7$ estimated from a small set of
EOSs by \cite{Lo2011}.
As we have pointed out in the derivation of Eq. \eqref{eqn:mtov_lo}, the
lower bound $M_{_{\rm TOV}}^{\min}$ on the maximum mass does not depend
on $\chi_2=\chi_{\rm Kep}$, due to the separate universality of the
maximum mass $M_{\rm crit} \left( \chi_\mathrm{Kep} \right)$ only in
terms of the $M_{_{\rm TOV}}$ (see also the discussion around
Eq. \eqref{eqn:M_crit}).  Thus, only the bounds on the spin $\chi_2$ will
explicitly depend on the precise value of $\chi_{\rm Kep}$ via
Eq. \eqref{eqn:chik}. For comparison we now adopt both the upper and
lower bounds on $C_{_{\rm TOV}}$ in terms of Eqs.  \eqref{eqn:C_TOV_max}
and \eqref{eqn:C_TOV_min}, which are consistent with GW170817
\citep{Most2020c}. In addition, we adopt very conservative ranges of $8
\leq R_{_{\rm TOV}}/{\rm km} \leq 15\ $, in order to determine the most
severe effect on the spin bound. The resulting spin bounds are shown in
Fig.  \ref{fig:appA}, which shows that the variations in the lower bound
of the secondary's spin are small and given by
\begin{align}
  0.49_{-0.05}^{+0.08} \lesssim \chi_2 \lesssim 0.68_{-0.05}^{+0.11}\,,
  \label{eqn:spinA}
\end{align}
under the assumption that $\MTOV \leq 2.3\,M_{\odot}$
\citep{Rezzolla2017,Shibata2019}. Stated differently, while it is useful
to consider an uncertainty in the value of $\chi_{\rm Kep}$, the latter
hardly affects our estimates \eqref{eqn:spinA} for the spin of the
secondary in GW190814.

\begin{figure}
  \centering \includegraphics[width=0.37\textwidth]{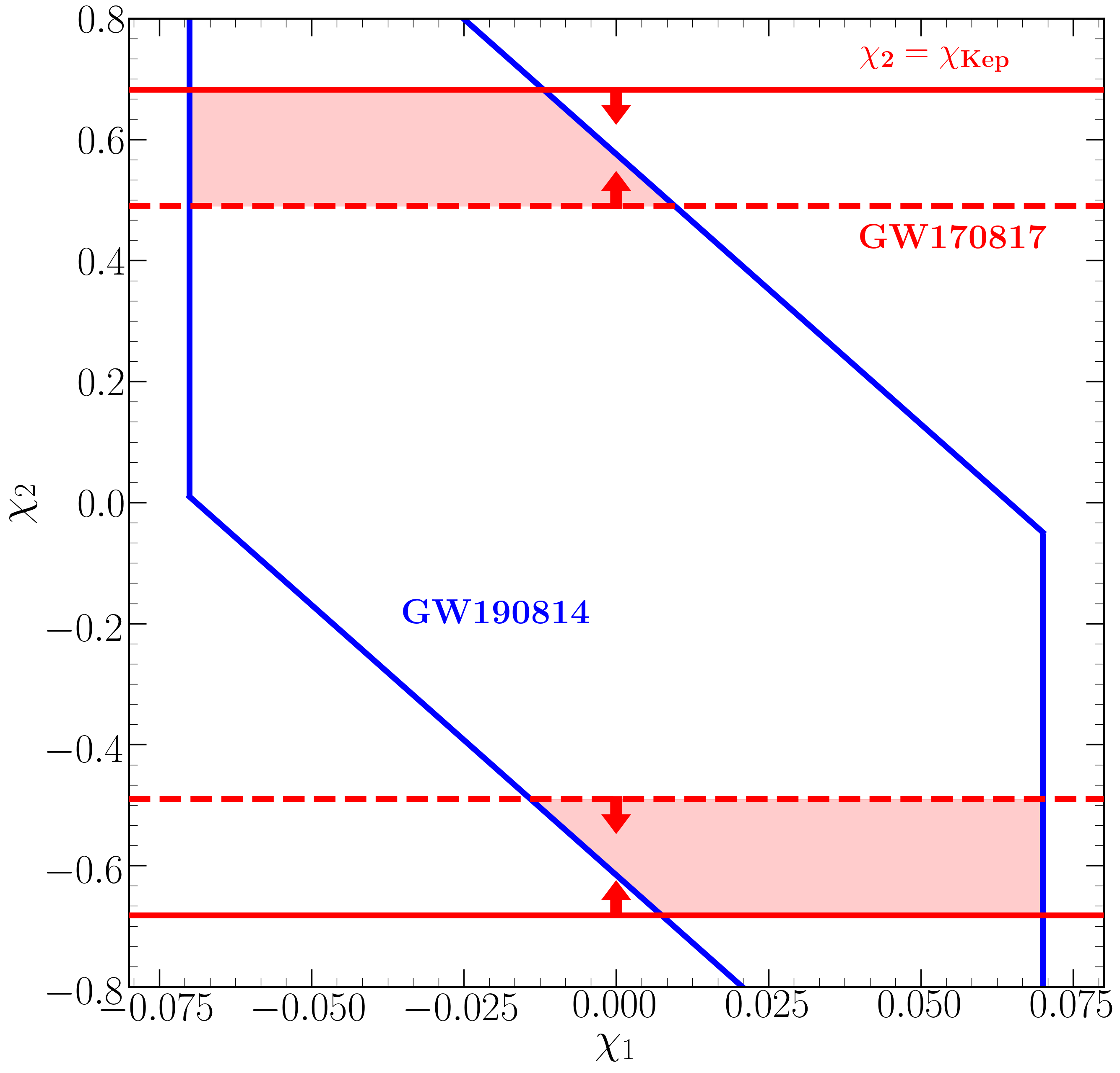}
  \caption{Allowed range (red-shaded area) of primary and secondary spins
    $\chi_1$ and $\chi_2$, in case the secondary was a NS at some point
    before merger.}
  \label{fig:chi12}
\end{figure}

\noindent Finally, this range for $\chi_2$ can be translated into a rotation
frequency of the NS, $\Omega_2=S_2/I_2$, that can be easily computed for a
given moment of inertia $I_2$, which can be expressed in terms of the NS
mass and radius. Using the fit for $\chi \approx 0.4$ from
\citet{Breu2016}, we can compute the moment of inertia as
\begin{equation}
  I_2/m_2^3 = \left(\bar{a}_1\mathcal{C}^{-1} +
  \bar{a}_2\mathcal{C}^{-2}
  +\bar{a}_3\mathcal{C}^{-3}
  +\bar{a}_4\mathcal{C}^{-4}
  \right)\, ,
\end{equation}
with the compactness $\mathcal{C} \coloneqq m_2/R_2$ and
$\bar{a}_1=9.50\times 10^{-1}$, $\bar{a}_2=1.44\times 10^{-2}$,
$\bar{a}_3=1.22\times 10^{-2}$, and $\bar{a}_4=-7.61\times 10^{-4}$.
Assuming a typical NS radius of $R_2=12.5\,(13\,\rm km)$ \citep{Most2018}
and $S_2=\chi_2 m_2^2$, we find a rotation frequency $f=\Omega/{2\pi}$ of
$1.21\,(1.14)\, \rm kHz$ for $\chi_2=0.49$. This frequency is
considerably higher than the fastest known pulsar PSR J1748--2446ad
\citep{Hessels2006}, with a frequency of $716\, \rm Hz$, thus making --
at least in this hypothetical scenario -- the secondary of GW190814 the
fastest known NS with a rotational kinetic energy $>10^{52}\,\rm erg$.
Assuming spins aligned/antialigned with the orbital angular momentum and
using the fit given in \citet{Barausse:2009uz} \citep[see
  also][]{Hofmann2016}, we can derive an estimate on the final BHs spin,
$\chi_{\rm fin}$. Assuming the spins of the primary and secondary are
anti-aligned, we derive $0.24 < \chi_{\rm fin} < 0.29$.
Finally, we check the consistency of our estimates by computing the
spin $\chi_1$ of the primary via Eq. \eqref{eqn:chitilde}
\begin{align}
  \chi_1 \left( \chi_2 \right) = - q \chi_2 + \tilde{\chi} \left( 1 + q \right),
\label{eqn:chi2}
\end{align}
and using $\tilde{\chi} = -0.002\pm 0.06$ and $q=0.112^{+0.008}_{-0.009}$
\citep{Abbott2020b}. We find that the values obtained here are all
consistent with those inferred by \citet{Abbott2020b}, \ie
$\left|\chi_1\right| < 0.07$. The overall allowed range of binary component
spins is then shown in Fig. \ref{fig:chi12}.

\section{Conclusion}

We have investigated how a lower bound on the maximum mass $\MTOV$ of a
nonrotating NS can be derived from the recent observation of the merger
of a $\sim 2.6\,M_{\odot}$ compact object with a $\sim 23\, \Msun$ BH
\citep{Abbott2020b}. More specifically, since the maximum-mass
constraints from GW170817 \citep{Margalit2017, Rezzolla2017, Ruiz2017,
  Shibata2019} and the observations of the very massive pulsars PSR 
J0348+0432 and PSR J0740+6620 \citep{Antoniadis2013,Cromartie2019}
indicate that the maximum mass of nonrotating NSs is lower than the
measured mass of the secondary $m_2$, rotation is needed to allow for the
secondary compact object in GW190814 to have been a NS at some point in
the inspiral. Using universal relations for the maximum mass of uniformly
rotating NSs \citep{Breu2016}, we infer a lower limit on the maximum mass
of nonrotating NSs, $\MTOV > 2.08^{+0.04}_{-0.04}\,M_{\odot}$. The new
lower limit on $M_{_{\rm TOV}}$ does not exclude EOSs supporting massive
nonrotating neutron stars, \ie $M_{_{\rm TOV}} \gtrsim 2.5\, M_\odot$,
although the studies that have explored this scenario have indicated that
$M_{_{\rm TOV}}$ cannot be much higher than the existing observational
limit \citep{Fattoyev2020}.

However, assuming the formation of a BH as the final remnant of GW170817,
makes it difficult to reconcile such high maximum masses with the amount
of angular momentum left after the merger and the ejected mass
\citep[see][for an extended discussion]{Gill2019,Shibata2019}. Hence,
imposing an upper limit $\MTOV \leq 2.3\,M_{\odot}$ consistent with the
multimessenger observation of GW170817 \citep{Rezzolla2017,Shibata2019},
restricts the amount of minimal spin of the secondary object to
$0.49_{-0.05}^{+0.08} \lesssim \chi_2 \lesssim 0.68_{-0.05}^{+0.11}$, a
quantity that has not been constrained by the observations of GW190814.
Conversely, if the maximum mass was as high as $2.51\, M_\odot$
\citep{Ruiz2017}, the spin required from Eq.  \eqref{eqn:M_crit} to
support a $2.6\, M_\odot$ NS, \ie the average secondary mass in GW190814,
would still be high, \ie $\chi_2 > 0.31$. Even for this high estimate for
the maximum mass, the rotation of the secondary can only be neglected, if 
the lower bound for the secondary mass, $m_2>2.51\, M_\odot$, is assumed 
\citep{Tsokaros2020}, in agreement with the allowed spin parameter space 
delimited by the blue lines in Fig. \ref{fig:chi12}.

\noindent 
Interestingly, since the rotational collapse of a magnetised
NS to a BH can be accompanied by the emission of a radio signal similar
to that measured in fast radio bursts (FRB) (\citet{Falcke2013} and also
\citet{Most2017}), there could be a potential connection between the
location of an FRB and of a massive binary merger of the type discussed 
here.
\vspace{-17pt}
\section*{Acknowledgements}
It is a pleasure to thank R. Essick, P. Landry, M. Safarzadeh, and
M. Zevin for useful discussions and comments.  Support comes in part from
HGS-HIRe for FAIR; the LOEWE-Program in HIC for FAIR; ``PHAROS'', COST
Action CA16214; the ERC Synergy Grant ``BlackHoleCam: Imaging the Event
Horizon of Black Holes'' (Grant No. 610058);

\vspace{-17pt}
\section*{Data availability}
No new data was generated or analysed in support of this research.
\vspace{-20pt}
\bibliographystyle{mnras}
\bibliography{aeireferences}

\end{document}